\documentclass[reprint,twocolumn,superscriptaddress,amsmath,amssymb,prb,nofootinbib]{revtex4}
\usepackage[dvipdfmx]{graphicx}
\usepackage{dcolumn}
\usepackage{bm}
\usepackage{color}
\usepackage{amssymb}
\usepackage{comment}
\usepackage{ulem}

\begin{document}

\title{Propagation effects in high-harmonic generation from dielectric thin films}

\author{Shunsuke Yamada}
\affiliation{Kansai Photon Science Institute, National Institutes for Quantum Science
and Technology (QST), Kyoto 619-0215, Japan}
\author{Tomohito Otobe}
\affiliation{Kansai Photon Science Institute, National Institutes for Quantum Science
and Technology (QST), Kyoto 619-0215, Japan}
\author{David Freeman}
\affiliation{Research School of Physics, Australian National University, 
Canberra, ACT, 2601, Australia}
\author{Anatoli Kheifets}
\affiliation{Research School of Physics, Australian National University, 
Canberra, ACT, 2601, Australia}
\author{Kazuhiro Yabana}
\affiliation{Center for Computational Sciences, University of Tsukuba, Tsukuba 305-8577, Japan}

\date{\today}

\begin{abstract}
Theoretical investigation is conducted of high-order harmonic
generation (HHG) in silicon thin films to elucidate the effect of
light propagation in reflected and transmitted waves.  The
first-principles simulations are performed of the process in which an
intense pulsed light irradiates silicon thin films up to 3~$\mu$m
thickness. Our simulations are carried within the time-dependent
density functional theory (TDDFT) with the account of coupled dynamics
of the electromagnetic fields and the electronic motion. It was found that
the intensity of transmission HHG gradually decreases with the thickness,
while the reflection HHG becomes constant from a certain thickness.
Detailed analyses show that transmission HHG have two origins: the HHG
generated near the front edge and propagating to the back surface,
and that generated near the back edge and emitted directly.  The
dominating mechanism of the transmission HHG is found to depend on the
thickness of the thin film and the frequency of the HHG.  At the film
thickness of 1~$\mu$m, the transmission HHG with the frequency below
20 eV is generated near the back edge, while that with the
frequency above 20~eV is generated near the front edge and
propagates from there to the back surface.
\end{abstract}
\maketitle

\section{Introduction}

The phenomenon of high-order harmonic generation (HHG) from solids irradiated with an intense ultrashort pulsed light has been very actively studied in recent years \cite{Ghimire2011, Schubert2014,  Vampa2015, Hohenleutner2015, Luu2015, Langer2017, You2017, Liu2017,  Shirai2018, Vampa2018, Orenstein2019, Ghimire2019, Li2020, Goulielmakis2022}.  
Apart from being at the focus of basic science, there is a strong interest from the perspective of applications, such as the creation of compact XUV light sources.  Numerous experimental investigations have been conducted on HHG in various  targets including simple dielectrics, two-dimensional materials \cite{Yoshikawa2017, Liu2017, Xia2018, LeBreton2018, Yoshikawa2019}, topological materials \cite{Bai2021, Heide2022}, and so on.  
Light pulses of linear, circular, or elliptical polarization have been used.  
Frequencies of light pulses range from terahertz to visible light.  
It has been shown that the HHG spectrum depends sensitively on the angle between the polarization direction of the light pulse and the crystal axes \cite{Langer2017, You2017, Neufeld2019}.

There have also been conducted numerous theoretical studies of HHG from solids examining the motion of electrons in a single unit cell of a crystal \cite{Yue2022}.  
Extensive discussions have been conducted on the mechanisms of HHG, either interband electronic transitions or intraband electronic motion.  Various theoretical approaches have been tried such as a one-dimensional model\cite{Hansen2017, Hansen2018, Jin2019, Ikemachi2017}, the time-dependent Schr\"odinger equation \cite{Wu2015, Apostolova2018}, density matrix models \cite{Ghimire2012, Vampa2014, Vampa2015-2}, the Floquet theory \cite{Faisal1997, Faisal2005}, and \textit{ab initio} descriptions such as the time-dependent density functional theory (TDDFT) \cite{Otobe2012, Otobe2016, Tancogne-Dejean2017, Floss2019, Tancogne-Dejean2022, Freeman2022}.

To explore HHG from solids, light propagation effects are another
important issue to be considered \cite{Orenstein2019, Hussain2021}.  
In the case of HHG from thin films, there appear higher-order harmonics in both transmitted and reflected waves \cite{Vampa2018-2, Xia2018}, which we call reflection HHG (RHHG) and transmission HHG (THHG), respectively, for which propagation effects work differently.  
There has also been interest in HHG from three-dimensional nanostructures to produce intense and characteristic HHG \cite{Vampa2017, Liu2018}.  
The precise way in which HHG depends on the shape of the nanostructure can only be investigated by examining the propagation effects.

Theoretical studies of HHG considering light propagation effects have
come to be actively studied recently \cite{Lorin2007, Ghimire2012, Floss2018, Orlando2019, Kilen2020, Yamada2021, Wu2022, Jensen2022}.  
For such studies, it is necessary to couple the description of electronic motion with the Maxwell equations which describe the light propagation. 

We previously reported the light propagation effect on RHHG and THHG from silicon films of thickness up to 200~nm \cite{Yamada2021}.
In that work, we developed and utilized the first-principles approaches combining the TDDFT for electronic motion with the Maxwell equations in two different schemes.  
For relatively thin (thick) films, calculations were performed using the single-scale\cite{Yamada2018} (multiscale\cite{Yabana2012}) Maxwell-TDDFT method in which the Maxwell and TDKS equations are coupled without (with) a coarse-graining approximation.  
It was found that the HHG signals are strongest in both the reflection and transmission waves at the film thickness of 2--15~nm.  
It was also found that there appears a clear interference effect  in the RHHG.

In the present work, we extend the first-principles Maxwell-TDDFT approach of the previous study to much thicker films up to 3~$\mu$m using the multiscale method with the course-graining approximation.
We suggest that it is important to clarify the propagation effect for materials whose size is comparable or larger than the wavelength of the laser pulse.  
Indeed, as will be shown, we find  two kinds of propagation effects for films of about 1~$\mu$m thickness. 
One is the nonlinear propagation effect of the intense incident pulse in the frequency region of the fundamental wave.  
Another is the propagation of the generated high harmonic wave that is described by the linear optical response in a wide spectral region.


This paper is organized as follows: Sec.~\ref{sec:method} describes
theoretical methods and numerical details.  In Sec.~\ref{sec:results},
the calculation results are presented and analyzed in detail.
Finally, a summary is presented in Sec.~\ref{sec:summary}.

\section{Theoretical methods \label{sec:method}}

\subsection{Single unit-cell calculation}

We first consider a simulation of the electronic motion in a unit cell of
a crystalline solid driven by a pulsed electric field ${\bm E}(t)$ with a
given time profile.  In the real time TDDFT, the electronic motion
is described by the following time-dependent Kohn-Sham (TDKS) equation
for the time-dependent Bloch orbitals $u_{n{\bm k}}({\bm r},t)$
\cite{Bertsch2000,Otobe2008},
\begin{equation}
    i\hbar \frac{\partial}{\partial t} u_{n{\bm k}}({\bm r},t)
    =
    h_{\rm KS}[{\bm A}(t)] u_{n{\bm k}}({\bm r},t),
    \label{eq:tdks}
\end{equation}
with the Hamiltonian
\begin{eqnarray}
h_{\rm KS}[{\bm A}(t)]    
    &=&  \frac{1}{2m} \left( -i\hbar \nabla + \hbar{\bm k} + \frac{e}{c} {\bm A}(t) \right)^2 - e\phi({\bm r},t) 
\nonumber\\    
    &+&  \hat v_{\rm{NL}}^{{{\bm k}+\frac{e}{\hbar c}{\bm A}(t)}} +
V_{\rm{xc}}({\bm r},t)\ . \quad
      \label{eq:hks}
\end{eqnarray}
Here we treat dynamics of the valence electrons with the norm-conserving
pseudopotential \cite{Troullier1991}.  The scalar potential,
$\phi({\bm r},t)$, is periodic and satisfies the equation
\begin{equation}
    \nabla^2 \phi({\bm r},t) = -4\pi e (n_{\rm ion}({\bm r}) - n_{\rm
      e}({\bm r},t) ) \ ,
\end{equation}
where ionic charge density $n_{\rm ion}({\bm r})$ is prepared so that
to produce the local part of the pseudopotential in $\phi$.  The
density of electrons, $n_{\rm e}({\bm r},t)$, is given by
\begin{equation}
    n_{\rm e}({\bm r},t) = \frac{1}{N_k}\sum_{n,{\bm k}}^{\rm occ}
    \vert u_{n{\bm k}}({\bm r},t) \vert^2,
\end{equation}
where $N_k$ is the number of $k$-points.
The sum is taken  over the occupied bands in the ground state.
The nonlocal part of the pseudopotential is modified as $\hat{v}_{{\rm NL}}^{{\bm k}}\equiv e^{-i{\bm k}\cdot{\bf r}}\hat{v}_{{\rm NL}}e^{i{\bm k}\cdot{\bm r}}$, where $\hat v_{\rm NL}$ is the usual separable form of the norm-conserving pseudopotential \cite{Kleinman1982}. 
$V_{\rm xc}({\bf r},t)$ is the exchange-correlation potential for which the adiabatic 
local-density approximation \cite{Perdew1981} is assumed.

The Bloch orbitals are initially set to the ground state solution.
Solving the TDKS equation, we obtain the electric current density
averaged over the unit-cell volume $\Omega$ as
\begin{eqnarray}
&&{\bf J}[{\bm A}(t)](t)=-\frac{e}{m}\int_{\Omega}\frac{d {\bm r}}{N_k \Omega}\sum_{n,{\bm k}}^{{\rm occ}}u_{n{\bm k}}^*({\bm r},t) \nonumber \\
&&\times \left\{-i\hbar\nabla+\hbar{\bm k}+\frac{e}{c}{\bm A}(t)+\frac{m}{i\hbar}\left[{\bm r},\hat{v}_{{\rm NL}}^{{{\bm k}+\frac{e}{\hbar c}{\bm A}(t)}}\right]\right\}u_{n{\bm k}}({\bm r},t),\nonumber \\
\label{eq:current}
\end{eqnarray}

Taking the Fourier transformation of the current, we obtain a spectrum
that will be used to analyze HHG,
\begin{equation}
    I(\omega) = \left\vert \int_0^{T_{\rm tot}} dt\, e^{i\omega
      t}f\left( \frac{t}{T_{\rm tot}} \right) {\bm J}[{\bm A}(t)](t)
    \right\vert^2 \ .
    \label{eq:spectrum_sc}
\end{equation}
Here we introduce a smoothing function $f(x) \equiv 1 - 3x^2 +
2x^3$. $T_{\rm tot}$ is the total calculation time.

\subsection{Multiscale Maxwell-TDDFT calculation}

We consider an irradiation of a free-standing thin film of thickness
$d$ in a vacuum by an ultrashort light pulse of a linearly polarized
plane wave at the normal incidence.  The multiscale Maxwell-TDDFT
method\cite{Yabana2012} is used to describe the light propagation.  In
the method, we combine the one-dimensional Maxwell equation that
describes the light propagation and a number of TDKS equations that
describe the electonic motion using a course-graining approximation.

The light propagation is described in the macroscopic scale by solving the
following wave equation:
\begin{equation}
 \left (\frac{1}{c^2} \frac{\partial^2}{\partial t^2}  - \frac{\partial^2}{\partial Z^2} \right) {\bm A}_{Z}(t)
= \frac{4\pi}{c}   {\bm J}_{Z}(t) \ ,
\label{eq:multiscale}
\end{equation}
where $Z$ is the macroscopic coordinate. ${\bm A}_{Z}(t)$ and ${\bm
  J}_{Z}(t)$ are the vector potential and the current density,
respectively.  This wave equation is solved using a one-dimensional
grid for the $Z$ variable.  At each grid point of $Z$ inside the film, we
consider an infinite crystalline system of the film material.  The
electronic motion at each grid point $Z$ is described by the Bloch
orbitals $u_{n{\bm k},Z}({\bm r},t)$ which satisfy the TDKS equation:
\begin{equation}
    i\hbar \frac{\partial}{\partial t} u_{n{\bm k},Z}({\bm r},t)
    =
    h_{\rm KS}[{\bm A}_Z(t)] u_{n{\bm k},Z}({\bm r},t) \ ,
    \label{eq:mtdks}
\end{equation}
where the Kohn-Sham Hamiltonian is given by Eq.~(\ref{eq:hks}).  Here,
the scalar potential and the exchange-correlation potential depend on
$Z$.  The macroscopic current density ${\bm J}_{Z}(t)$ at each $Z$ is
expressed using Eq.~(\ref{eq:current}) as
\begin{equation}
    {\bm J}_Z(t) = {\bm J}[{\bm A}_Z(t)](t) \ .
    \label{eq:mcurrent}
\end{equation}

By solvingy Eqs. (\ref{eq:multiscale}), (\ref{eq:mtdks}) and
(\ref{eq:mcurrent}) simultaneously, we obtain the solution ${\bm
  A}_Z(t)$ for a given initial pulse that is prepared in the vacuum
region in front of the thin film.  At the beginning of the
calculation, the Bloch orbitals $u_{n {\bm k},Z}({\bm r},t)$ are set
to those of the ground state for all grid points of $Z$.

The multiscale Maxwell-TDDFT calculation provides the time profile of
the electric field, ${\bm E}_Z(t)$.  The time profiles of the
reflected and transmitted fields are given by those at the grid points
of the front ($Z=0$) and back ($Z=d$) edges, respectively, where
the incident pulse is subtracted for the former.  We evaluate the HHG
spectra via the square of the Fourier-transformed electric field
$|\tilde{E}(\omega)|^2$, where $\tilde{E}(\omega)$ is defined by
\begin{equation}
   \tilde{E}(\omega)= \int_{0}^{ T_{\rm tot}} dt \, e^{i\omega t} E(t)
   \, f\left(\frac{t}{T_{\rm tot}}\right) \ .
   \label{eq:fourier}
\end{equation}
Here, $T_{\rm tot}$ is the total calculation time for which we use
$T_{\rm tot}=200$ fs.  In Eq.~(\ref{eq:fourier}), we use again a
smoothing function $f(x)\equiv 1-3 x^2 +2 x^3$ to remove noises that
originate from the end of the pulse.  We further apply a frequency
averaging using the Gaussian convolution as follows:
\begin{equation}
\langle |\tilde{E}(\omega)|^2 \rangle = \int d\omega' \frac{e^{-
    (\omega-\omega')^2/\epsilon^2}}{\sqrt{\pi \epsilon^2}}
|\tilde{E}(\omega')|^2 \ .
\end{equation}
Here we will use $\epsilon=5\times 10^{-3}$ a.u.  When we examine
strength of the $n$th-order harmonics, we will use:
\begin{equation}
    I_n = \int^{\left(n+\frac{1}{2}\right)\omega_0}_{\left(n-\frac{1}{2}\right)\omega_0} d\omega \,
    |\tilde{E}(\omega)|^2 \ ,
    \label{eq:harmonics}
\end{equation}
where $\omega_0$ is the frequency of the fundamental wave.

\subsection{Numerical detail}

We utilize an open-source software package SALMON (Scalable Ab initio
Light-Matter simulator for Optics and
Nanoscience)\cite{Noda2019,SALMON_web} for which some of the present
authors are among the leading developers.  In this code, solutions for
the electronic orbitals as well as the electromagnetic fields are found by
using the finite-difference method.  The time evolution of the
electron orbitals is carried out using the Taylor expansion method
\cite{Yabana1996}.

In solving the TDKS equation, we use a conventional unit cell of the diamond structure containing eight Si atoms with the lattice constant of $a = 0.543$ nm.
The numbers of grid points for discretizing the unit cell volume and the Brillouin zone are set to $N_r=16^3$ and $N_k=32^3$, respectively.
A uniform 1D grid is introduced to describe the wave equation of Eq.~(\ref{eq:multiscale})
with the grid spacing of $6.25$ nm, except the $d=5$~nm case where the grid spacing of $5$ nm is used, that is, the film is expressed as a single grid point.
The time step is set to $2.5 \times 10^{-3}$ fs.
We have carefully examined the convergence of the calculations with respect to the discretizations in spatial, $k$-space, and time variables.

\section{Results and discussion \label{sec:results}}

\subsection{Single unit-cell calculation}

\begin{figure}
    \includegraphics[keepaspectratio,width=\columnwidth]
    {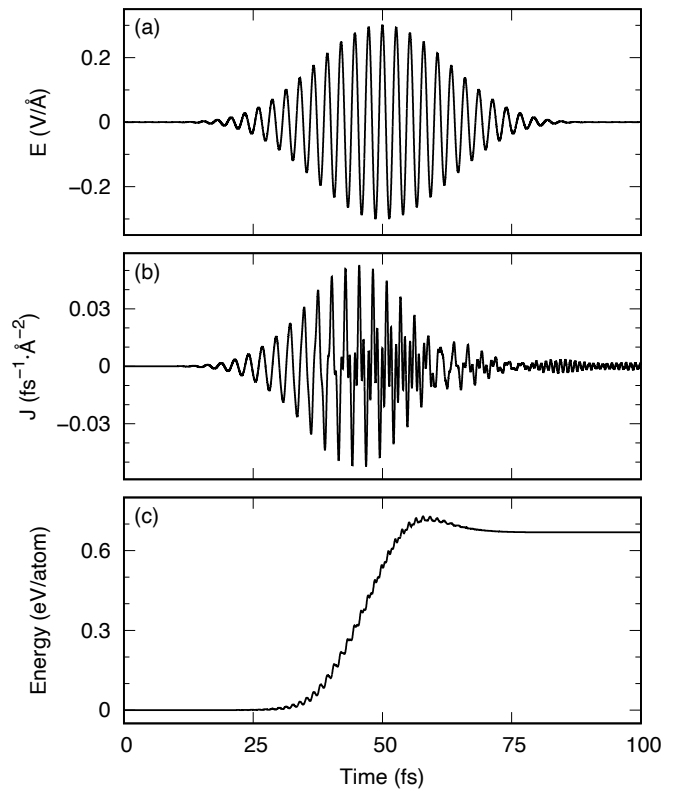}
    \caption{\label{fig:sc_time} 
    (a) Time profile of the applied electric field with the maximum amplitude of $E_0=0.3$~V/{\AA}. (b) The induced current density in the unit cell of Si. (c) The energy deposition per atom to electrons in the unit cell of Si.
    }
\end{figure}

\begin{figure}
    \includegraphics[keepaspectratio,width=\columnwidth]
    {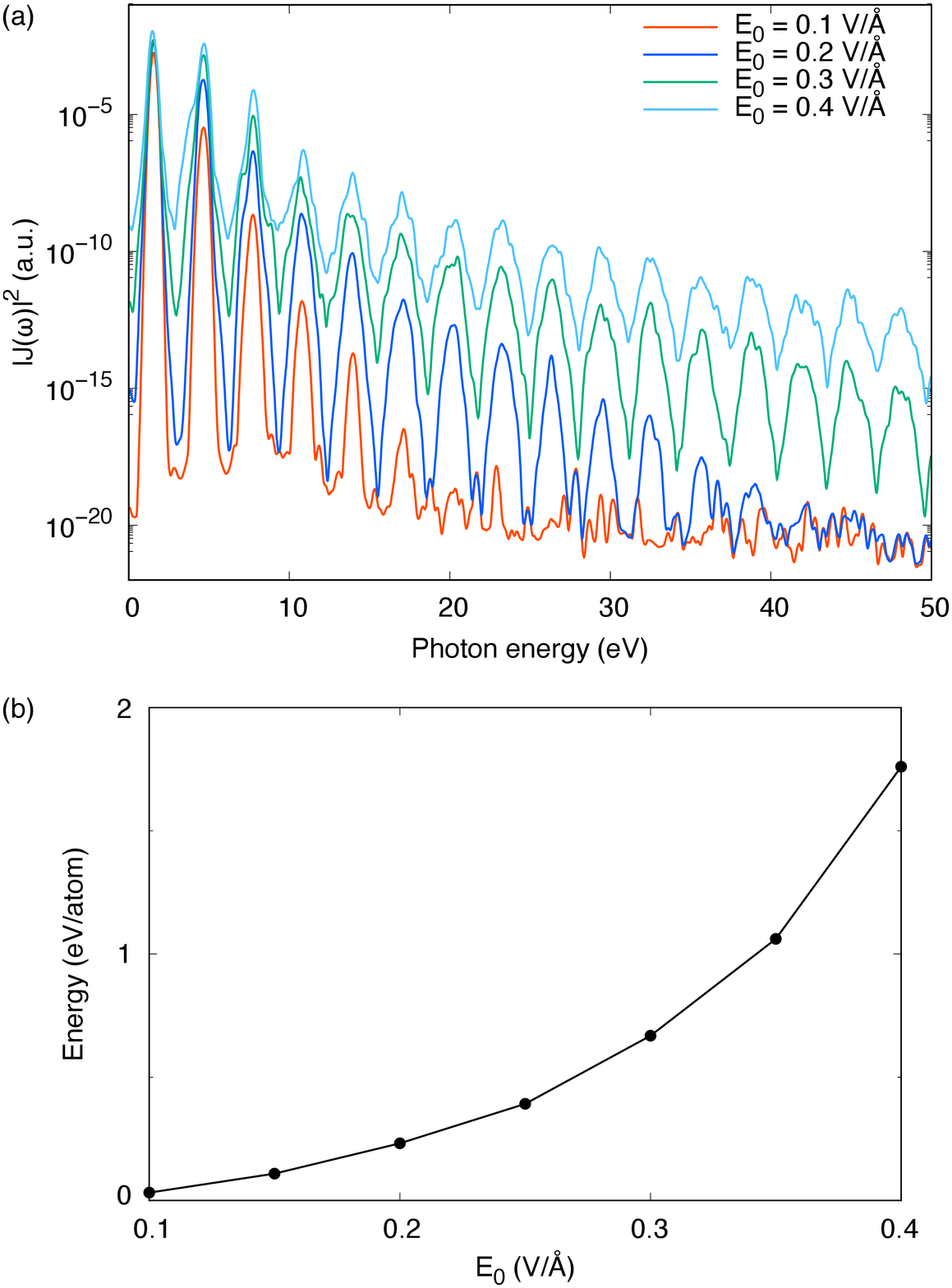}
    \caption{\label{fig:sc} (a) HHG spectra calculated in the unit
      cell of Si for an applied electric field of
      Eq.~(\ref{eq:incident}) with the maximum amplitude of $E_0=0.1$,
      $0.2$, $0.3$, and $0.4$~V/{\AA}.  (b) The energy deposition per
      atom after the pulse end plotted against the maximum amplitude
      of the applied electric field.  }
\end{figure}

Before discussing propagation effects in HHG, we show calculations of electronic motion and HHG spectrum in a unit cell of crystalline Si under a pulsed electric field of a given time profile. 
They will be useful to understand results of multiscale Maxwell-TDDFT calculations that involve complex coupling with light propagation.

We employ a pulsed electric field of linear polarization given by the following time profile for the vector potential,
\begin{eqnarray}
{\bm A}(t) = - \frac{c E_0}{\omega_0} \, \sin \left[ \omega_0 \left( t
  - \frac{T}{2} \right) \right] \,\, \sin^6 \left( \frac{\pi t}{T}
\right) \hat{\bm x} , \nonumber \\ (0<t<T).
\label{eq:incident}
\end{eqnarray}
Here $E_0$ is the maximum amplitude of the electric field, $\omega_0$
is the fundamental frequency, and $T$ is the total duration of the pulse.
The value of the maximum amplitude $E_0$ will be specified later.  In
the following calculations, we set as $\hbar \omega_0=$ 1.55 eV and
$T=$ 100 fs.  The electric field ${\bm E}(t)$ is related to the vector
potential ${\bm A}(t)$ by ${\bm E}(t) = - (1/c)(d{\bm A}/dt)$.

In Fig.~\ref{fig:sc_time}, we show results of a typical calculation
solving the TKDS equation~(\ref{eq:tdks}).  Panel (a) shows the
applied pulse shape of Eq.~(\ref{eq:incident}) with the maximum
amplitude of $E_0=0.3$ V/{\AA}.  Panel (b) shows the induced current
density averaged over the unit cell volume given by
Eq.~(\ref{eq:current}).  At the beginning of the calculation, the
induced current looks proportional to the applied electric field.  In
fact, there is a phase shift of $\pi/2$ since the average frequency of
the pulse is below the direct bandgap of Si that is equal to
2.6 eV in the local density approximation.  At
around the peak of the applied electric field and immediately after,
there appear high-frequency oscillations in the current that are
related to HHG.  Panel (c) shows the electronic excitation energy per
atom.  Although the average frequency is below the bandgap energy, we
find an increase of electronic excitation energy due to nonlinear
excitation.  The amount of the excitation energy after the applied
electric field termination is about 0.68 eV per atom.  We consider that this
amount of the excitation energy is close to, but still below, the value
that causes permanent damage to a surface of bulk Si\cite{Bonse2002,
  Medvedev2015,Venkat2022,Venkat2022a}.

Figure~\ref{fig:sc}(a) shows HHG spectra of single-cell calculations, Eq.~(\ref{eq:spectrum_sc}), for applied pulses of several intensities, $E_0=0.1$, $0.2$, $0.3$, and $0.4$ V/{\AA}. 
The average frequency and the pulse duration are chosen to be common as $\hbar\omega$ = 1.55 eV and $T=100$ fs, respectively.
As is evident from the figure, the HHG becomes more pronounced and extended to higher orders as the field amplitude increases. 
Clear HHG signals are seen up to 11th order for $E_0=0.1$ V/{\AA}, up to 25th for $E_0=0.2$ V/{\AA}, and more than 30th for $E_0=0.3$ V/{\AA}. We note that production of HHG of very high order can be observed by using sufficiently long pulse, as we have shown previously \cite{Freeman2022}.

Figure~\ref{fig:sc}(b) shows the energy deposition to electrons after
the pulse end.  Since two photons are required to excite electrons
across the direct bandgap, the energy deposition should scale as
$\Delta E \propto E_0^4$ at low intensity region.  For the pulse of
the maximum amplitude of 0.35 eV/{\AA}, the energy deposition per atom
exceeds 1.0 eV per atom that eventually leads, as we mentioned
previously, to a permanent damage to the material.  These results
indicate that a pulse of maximum amplitude around 0.3~V/{\AA} produces
HHG of very high orders extending beyond 50~eV, while avoiding to produce
permanent damage to the material.

\subsection{Multiscale calculation}

\subsubsection{Light propagation}

\begin{figure}
    \includegraphics[keepaspectratio,width=\columnwidth]
    {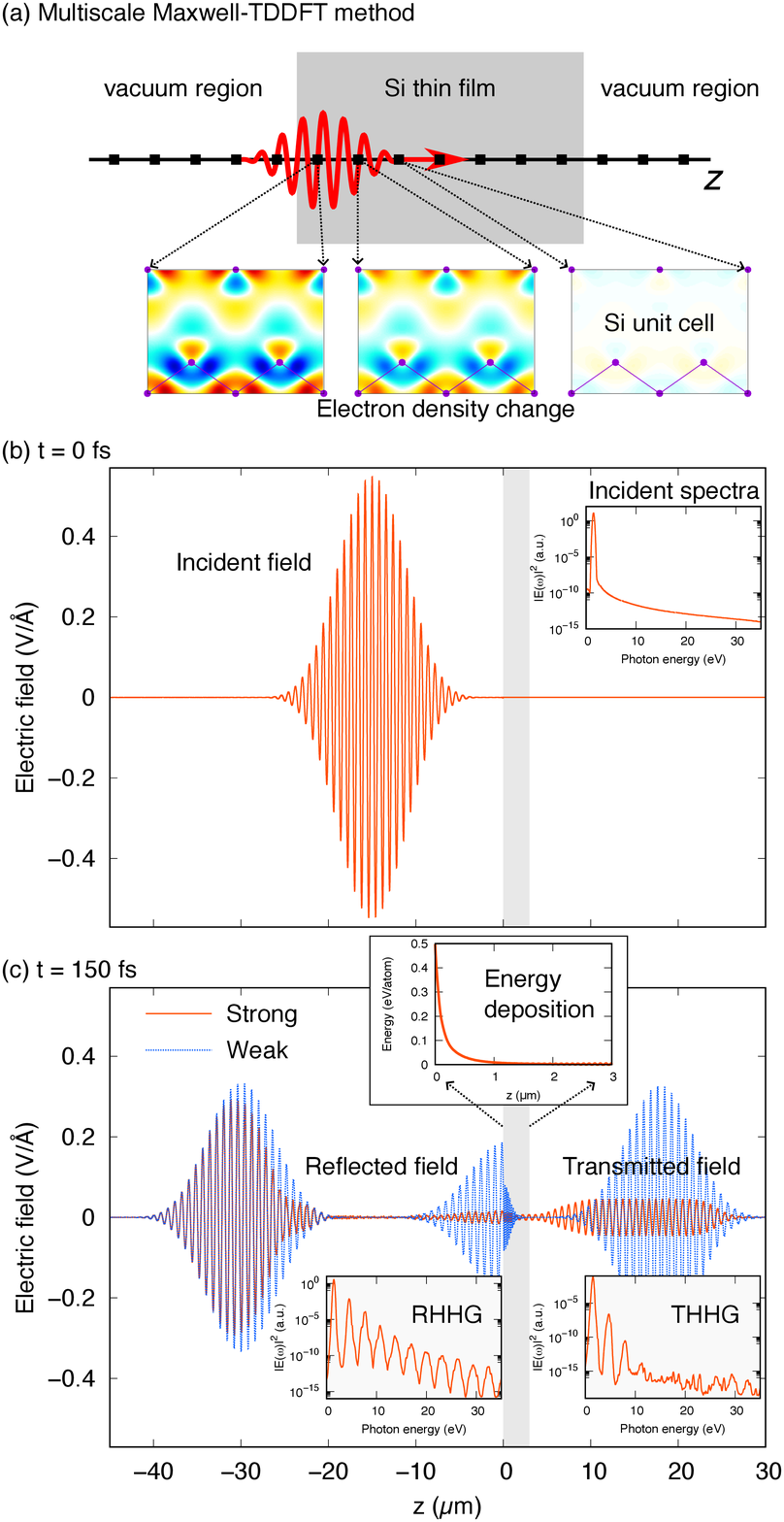}
    \caption{\label{fig:pulse} (a) Overview of the multiscale
      Maxwell-TDDFT method for a light propagation through a Si thin
      film. The electron density changes driven by the light pulse are
      illustrated for the first three grid points.  (b) Electric field
      at $t=0$ is shown. In front of the Si thin film that is exhibited as
      a gray area, the incident pulse is prepared.  In the inset, a
      Fourier spectrum of the incident electric field is shown.  (c)
      Electric field at $t=150$ fs is shown for the case of two
      incident pulses: a strong pulse ($I=4\times 10^{12}$ W/cm${}^2$,
      red solid line) and a weak pulse ($I= 10^{9}$ W/cm${}^2$, blue
      dotted line) scaled up by a factor of $\sqrt{4000}$.  In the
      upper inset, the energy deposition as a function of penetration
      depth is shown.  In the lower insets, a Fourier spectrum of the
      reflected and transmitted pulses are shown for the case of a
      strong incident pulse.  }
\end{figure}

Figure~\ref{fig:pulse} summarizes an overview of the multiscale Maxwell-TDDFT calculation for the light propagation of a pulsed light through a Si thin film.
Figure~\ref{fig:pulse}(a) shows a schematic view of the numerical method.
The light propagation is solved on a uniform 1D grid along the $Z$-axis. 
At each grid point inside the thin film, electron dynamics calculation is carried out in a unit cell of Si.
In the figure, electron density changes driven by the light pulse are illustrated for the first three grid points of the $Z$-coordinate.

Figure~\ref{fig:pulse}(b) and (c) show snapshots of the electric field
of the light pulse propagating through a Si thin film of $d=3000$ nm
thickness.  In Fig.~\ref{fig:pulse}(b), the electric field of the
initial pulse ($t=0$ fs) that locates in front of the film is shown
where the film is exhibited as a thin gray area.  The initial pulse
with the time profile given by Eq.~(\ref{eq:incident}) is used with
the fundamental frequency of $\hbar\omega=1.55$ eV, total duration of
$T=100$ fs, and the maximum intensity of $I=4 \times 10^{12}$~W/cm$^2$
that corresponds to $E_0 = 0.55 $~V/{\AA}.  Note that the maximum
field amplitude and the maximum intensity is related by
$I=cE_0^2/(8\pi)$.  In Fig.~\ref{fig:pulse}(c), a snapshot of the
electric field at $t=150$ fs is shown.  Here we display results
corresponding to initial pulses of two different maximum intensities,
$I=4 \times 10^{12}$~W/cm$^2$ by red-solid line and $I=1.0 \times
10^9$ W/cm$^2$ by blue-dotted line, respectively.  For the weaker
pulse of the initial intensity of $I=1.0 \times 10^9$~W/cm$^2$, linear
propagation is expected.  In the figure, the field is multiplied by a
factor of $\sqrt{4000}$ so that the differences of two lines manifest
nonlinear effects in the stronger pulse.  In the snapshot, reflected
and transmitted pulses are seen. They are apart from the film (left
for the reflected and right for the transmitted pulses).  There also
appears a component around the film that is caused by a reflection at
the back surface of the film.

It can be observed that the nonlinear effects appear more significantly in the transmitted pulse than in the reflected pulse.
While the envelopes of the reflected pulses of the weaker and the stronger cases look similar to each other and do not change much from the envelope of the initial pulse, the transmitted pulse of the stronger case suffers a strong nonlinear effect with the nearly flat envelope.
This can be understood as follows: During the propagation, the high field component of the electric field excites efficiently electrons in the medium and, as the reaction, the amplitude of the high field part is reduced to produce the flat envelope in the transmitted pulse.

As the inset of Fig.~\ref{fig:pulse}(c), the energy deposition from
the light field to electrons in the unit cell is shown as a function
of $Z$.  At the front surface, the energy deposition is about
0.5~eV/atom and is close to the case of 0.25~V/{\AA} pulse in the
single unit-cell calculation shown in Fig.~\ref{fig:sc}(b).  Since the
electric field at the front surface is given as the sum of those of the
incident and the reflected pulses, the maximum electric field at the
front surface is smaller than that of the incident pulse that is more
than 0.5~V/{\AA}.  As we noted previously, this amount of the energy
deposition will not cause a permanent damage to the material.  The
fluence of the incident pulse is about 0.09~J/cm$^2$ and this value is
substantially smaller than 0.2~J/cm$^2$, which is the reported
threshold value of Si modification\cite{Bonse2002}.  The energy
deposition at the front surface, 0.5~eV/atom, is also smaller than
0.65~eV/atom, which is the theoretical value of the threshold for the
phase transition\cite{Medvedev2015}.  The energy deposition shows the
maximum value at the front surface and decays quickly inside the
material, less than 0.1 eV/atom at 200~nm from the surface.  This is
because the high amplitude component of the light pulse is reduced
during the propagation by nonlinear interaction.

\subsubsection{HHG spectra}

A Fourier spectrum [Eq.~(\ref{eq:fourier})] of the incident pulse is
shown in the inset of Fig.~\ref{fig:pulse}(b), and those of the
reflected and transmitted pulses (RHHG and THHG) are shown in the
insets of Fig.~\ref{fig:pulse}(c) for the case of a strong incident
pulse of $I=4.0 \times 10^{12}$ W/cm$^2$.  We note that all the
Fourier transforms are taken with a sufficiently long duration of
$T_{\rm tot}=200$ fs, in which the pulse reflected at the back surface
is included in the RHHG.  It can be seen that, although both the
reflected and transmitted pulses include HHG components, RHHG shows
clear signals of higher orders than THHG.

\begin{figure}
    \includegraphics[keepaspectratio,width=\columnwidth]
    {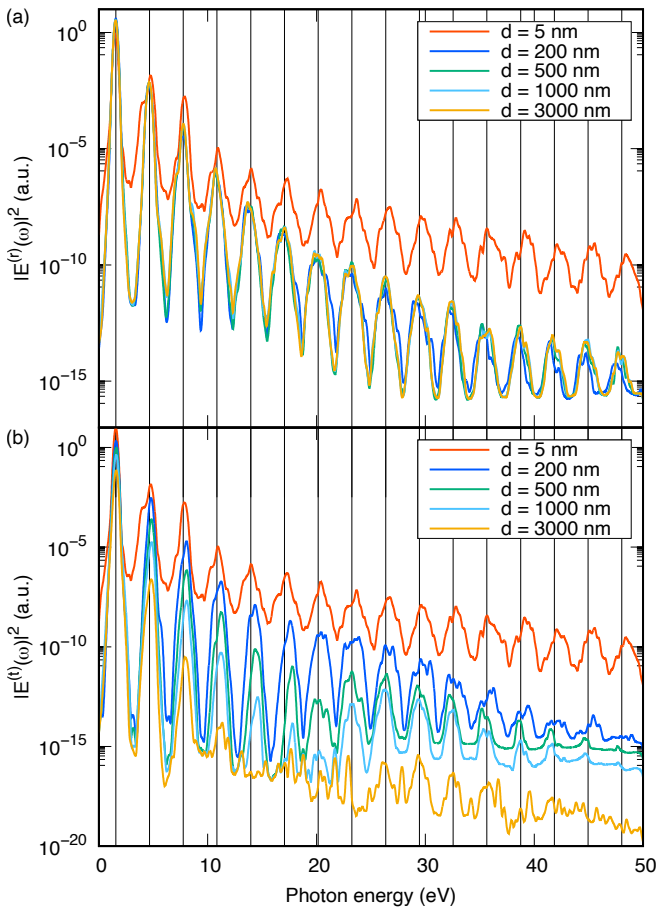}
    \caption{\label{fig:hhg} 
    Spectra of RHHG (a) and THHG (b) are shown for films of thickness $d$ of 5, 200, 500, 1000, and 3000 nm.
    }
\end{figure}

\begin{figure}
    \includegraphics[keepaspectratio,width=\columnwidth]
    {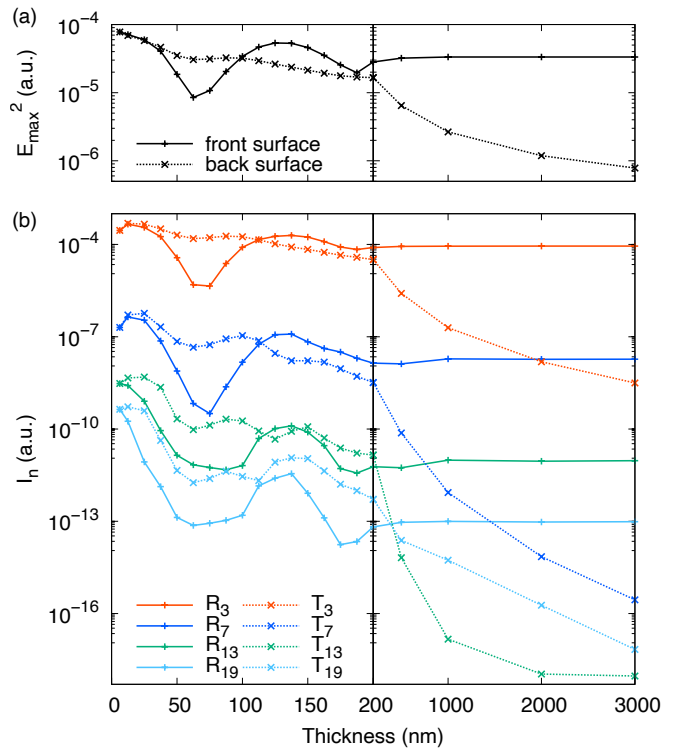}
    \caption{\label{fig:thickness} 
    (a) Square of the maximum electric field amplitude at the front surface (solid line) and back  surface (dotted line) of the film is shown for films of various thickness.
    (b) Strengths [Eq.~(\ref{eq:harmonics})] of the 3rd, 7th, 13th, and 19th harmonics for RHHG (solid lines) and THHG (dotted lines) are shown for films of various thickness.
    }
\end{figure}

We next examine thickness dependence of RHHG and THHG.
Figure~\ref{fig:hhg}(a) and (b) show spectra of RHHG and THHG, respectively, for Si films with thickness $d$ of 5, 200, 500, 1000, and 3000 nm.
The incident pulse is chosen to be the same as that shown in Fig.~\ref{fig:pulse}(b).
Figure~\ref{fig:thickness} shows thickness dependence of RHHG and THHG at several harmonics orders for films of various thickness up to 3000 nm. 
In Fig.~\ref{fig:thickness}(b), strengths of RHHG (solid lines) and THHG (dotted lines) given by Eq.~(\ref{eq:harmonics}) are shown for the 3rd (4.65 eV), 7th (10.85 eV), 13th (20.15 eV), and 19th (29.45 eV) harmonics.
In Fig.~\ref{fig:thickness}(a), a square of the maximum amplitude of the electric field in time, $E_{\rm max}=\max_t |E(t)|$, are shown at the front surface (solid line) and back surface (dotted line) of the film.
We note that a similar plot was provided in Fig.~5 of Ref~\onlinecite{Yamada2021} up to 200 nm thickness. 

As is seen from Fig.~\ref{fig:hhg}(a) and (b) and Fig.~\ref{fig:thickness}(b), spectra of RHHG and THHG for the film thickness of 5 nm are almost identical with each other and are the strongest among signals of films of different thicknesses. 
These features have already been reported in our previous analysis \cite{Yamada2021} and can be understood in terms of two-dimensional approximation for electromagnetism that is valid for very thin films.

In Fig.~\ref{fig:thickness}(b), there appears an oscillatory behavior
as a function of thickness, clearly in RHHG and vaguely in THHG, below
200 nm. It originates from the interference effect that is seen in
Fig.~\ref{fig:thickness}(a) at the front surface \cite{Yamada2021}.
Since the average frequency of the incident pulse, $\hbar\omega=1.55$
eV is below the direct bandgap of Si that is 2.6
eV in the present LDA calculation, the propagation is regarded as in a
transparent medium in linear optics.  Then, we expect that the
electric field at the front surface suffers a cancellation if the
thickness is equal to $(\lambda/n)(2m+1)/2$, where $\lambda$ is the
wavelength of incident light, $n$ is the index of refraction, and $m$
is an integer.  In Fig.~\ref{fig:thickness}(a), the maximum amplitude
shows a dip at the thickness of $d \sim 50$ nm. This corresponds to
$m=0$ case using the value of index of refraction, $n \sim 4$.
However, such an interference effect soon becomes ineffective as the
thickness increases since the strong pulse attenuates quickly during
the propagation due to nonlinear excitation processes.  As seen in
Fig.~\ref{fig:thickness}(a), the next cancellation expected at around
$d = 150 \sim 200$ nm is less clear.  We should note that the
thickness at which the interference disappears depends strongly on the
choice of the intensity of the initial pulse.  In the present setting,
we consider that the interference is no more important for films
thicker than 200 nm.

In both Figs.~\ref{fig:hhg}(b) and ~\ref{fig:thickness}(b), as the
thickness increases from 200 nm to 3000 nm, the RHHG signals change
little by the thickness. Meanwhile, the THHG signals decrease
substantially and monotonically with the thickness increase.  These
trends look to follow the thickness dependence of the maximum
amplitude of the electric field at the front and the back surfaces
shown in Fig.~\ref{fig:hhg}(a).  This finding indicates that RHHG
(THHG) is dominantly produced at the front (back) surfaces. However,
there takes place more complex dynamics that can be found if we look
carefully the thickness dependence of THHG.

We focus on the THHG generated from the film of thickness $d=1000$ nm
that is shown by light blue line in Figs.~\ref{fig:hhg}(b).  At this
thickness, we find an appearance of a dip in THHG at around 20 eV.
Looking at the spectrum carefully, peaks of THHG below 20 eV show blue
shift, while peaks above 20 eV do not.  We also note that THHG
spectrum for the film of $d=3000$ nm thickness has no longer clear
peaks except for a few peaks at low frequency region.  To understand
the mechanism that causes these features, we make an analysis
decomposing propagating pulse into frequency window in the next
subsection.

\subsubsection{Analysis using frequency windows}

\begin{figure*}
    \includegraphics[keepaspectratio,width=\textwidth]
    {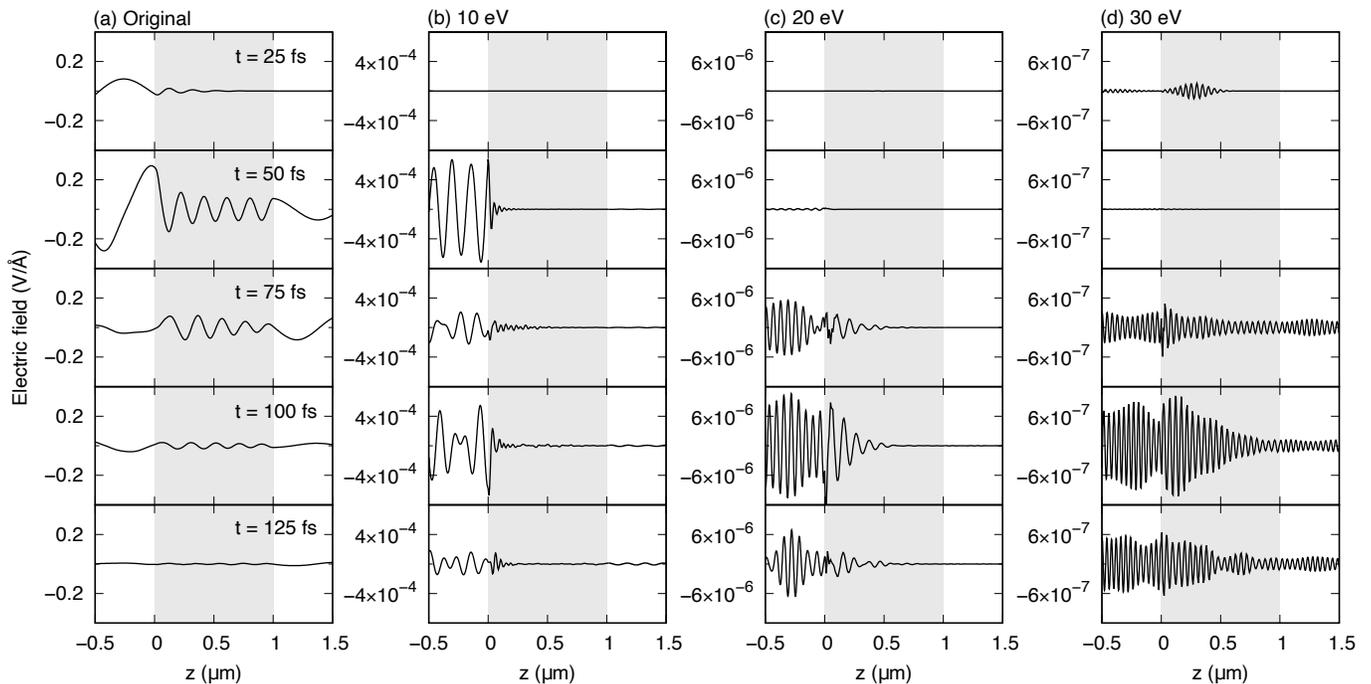}
    \caption{\label{fig:pulse_inv} 
    (a) Snapshots of the electric field at $t=25$, $50$, $75$, $100$, and $125$ fs for the pulse propagation through the Si film of $d=1000$ nm thickness.
    (b,c,d) The same as (a) but for the filtered inverse-Fourier transforms [Eq.~(\ref{eq:inverse})]  using the Blackman window with the central frequency of 10, 20, and 30 eV, respectively.
    }
\end{figure*}

To understand mechanisms that produce characteristic features of THHG
mentioned above, we investigate the pulse propagation behavior
separating frequency regions using a filtered inverse-Fourier
transformation as described below.  Once the multiscale Maxwell-TDDFT
calculation is finished, we take a Fourier transformation to obtain
the electric field in the frequency domain at each grid point $Z$,
$E_Z(\omega)$.  We then perform the inverse-Fourier transformation
with the frequency window as follows,
\begin{equation}
  E^{\rm w}_{Z}(t)= \int_0^{\omega_{\rm max}} \frac{d\omega }{\pi} {\rm Re} \left[
  e^{-i\omega t} E_Z(\omega) w(\omega)
  \right].
  \label{eq:inverse}
\end{equation}
As the window function for filtering the frequency region,
$w(\omega)$, we use the Blackman window function,
\begin{equation}
    w(\omega)=w^{}_{\rm Blackman} \left( \frac{\omega-\omega'}{\omega_{\rm width}/2} \right),
\end{equation}
where $\omega'$ is the central frequency and the frequency width $\omega_{\rm width}$ is set as $\hbar \omega_{\rm width}=10$ eV.

Figure~\ref{fig:pulse_inv}(a) shows snapshots of the electric field at
$t=25$, $50$, $75$, $100$, and $125$ fs for a pulse propagation
through the Si thin film of $d=1000$ nm thickness.  The gray area
indicates the Si thin film.  We note that the incident pulse ends at
$t=100$ fs.  Figure~\ref{fig:pulse_inv}(b), (c), and (d) show
frequency-gated snapshots of the propagating pulse using the filtered
inverse-Fourier transform of Eq.~(\ref{eq:inverse}) with the central
frequency of $\hbar\omega'= 10$, 20, and 30~eV, respectively.

We first look at Fig.~\ref{fig:pulse_inv}(a) which shows snapshots of the whole pulse.
The amplitude is maximal at $t=50$~fs at the front surface.
Since HHG depends strongly on the amplitude of the electric field as shown in Fig.~\ref{fig:sc}(a), HHG takes place most efficiently at this moment and at around the front surface.

We next look at $\hbar\omega'=10$ eV case shown in
Fig.~\ref{fig:pulse_inv}(b).  There appears a strong reflection wave
that corresponds to the RHHG around 10 eV which is produced at the
front surface.  However, inside the medium, the wave generated at the
front surface attenuates quickly as it propagates.  This is because
silicon is strongly absorptive at around 10 eV in linear optics.
Nevertheless, we find a weak transmitted wave at $t=100$ and 125 fs
that should correspond to the THHG around 10 eV.  This THHG should be
produced at the back surface of the film.  As seen in
Fig.~\ref{fig:pulse_inv}(a) at $t=50$ fs, there is an electric field
of amplitude about 0.1~V/{\AA} at around the back surface that can
produce HHG up to 20~eV, as shown in Fig.~\ref{fig:sc}.

The waveform looks similar at 20~eV region, as shown in
Fig.~\ref{fig:pulse_inv}(c), except for one important
difference. There appears no visible transmission wave at $t=100$ and
125~fs.  This can be understood by noting that the electric field at
the back surface, shown in Fig.~\ref{fig:pulse_inv}(a) at $t=50$~fs,
is not large enough to produce HHG around 20~eV.

In contrast, at 30 eV case shown in Fig.~\ref{fig:pulse_inv}(d), the HHG wave produced at the front surface does not attenuate completely and appears as the transmitted wave. 
It indicates that THHG at 30 eV region is mainly produced at the front surface and propagate through the film.

\begin{figure}
    \includegraphics[keepaspectratio,width=\columnwidth]
    {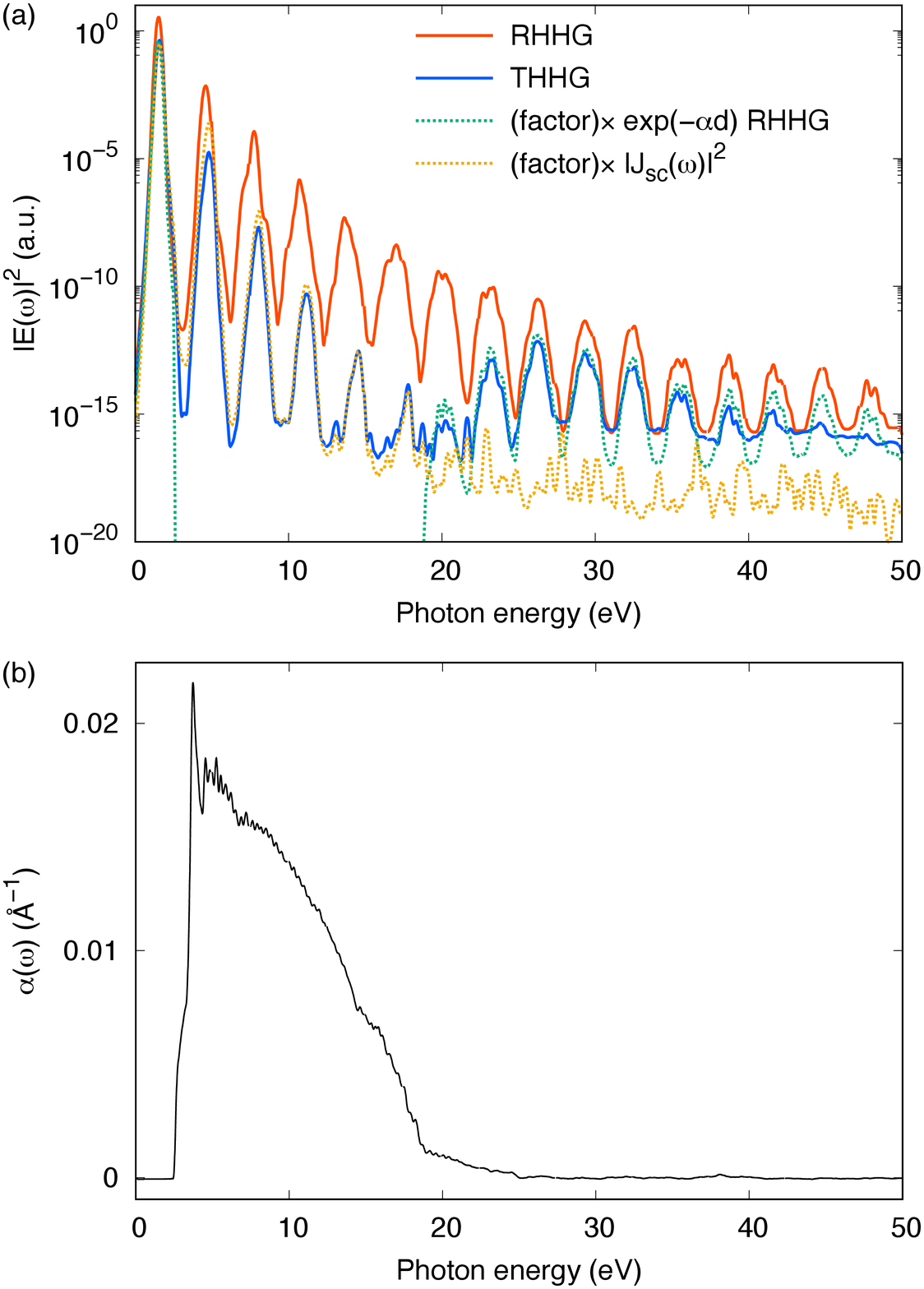}
    \caption{\label{fig:hhg_d1000} 
    (a) HHG spectra for the $d=1000$ nm case. The red and blue solid lines are RHHG and THHG, respectively.
    The green dotted line is RHHG multiplied by a constant factor and the term of $\exp(-\alpha(\omega)d)$, where $\alpha(\omega)$ is the attenuation coefficient of Si.
    The orange dotted line is a spectrum of the electric current in the Si unit cell driven by the fundamental wave at the back surface.
    (b) Attenuation coefficient $\alpha(\omega)$.
    }
\end{figure}

\begin{figure}
    \includegraphics[keepaspectratio,width=\columnwidth]
    {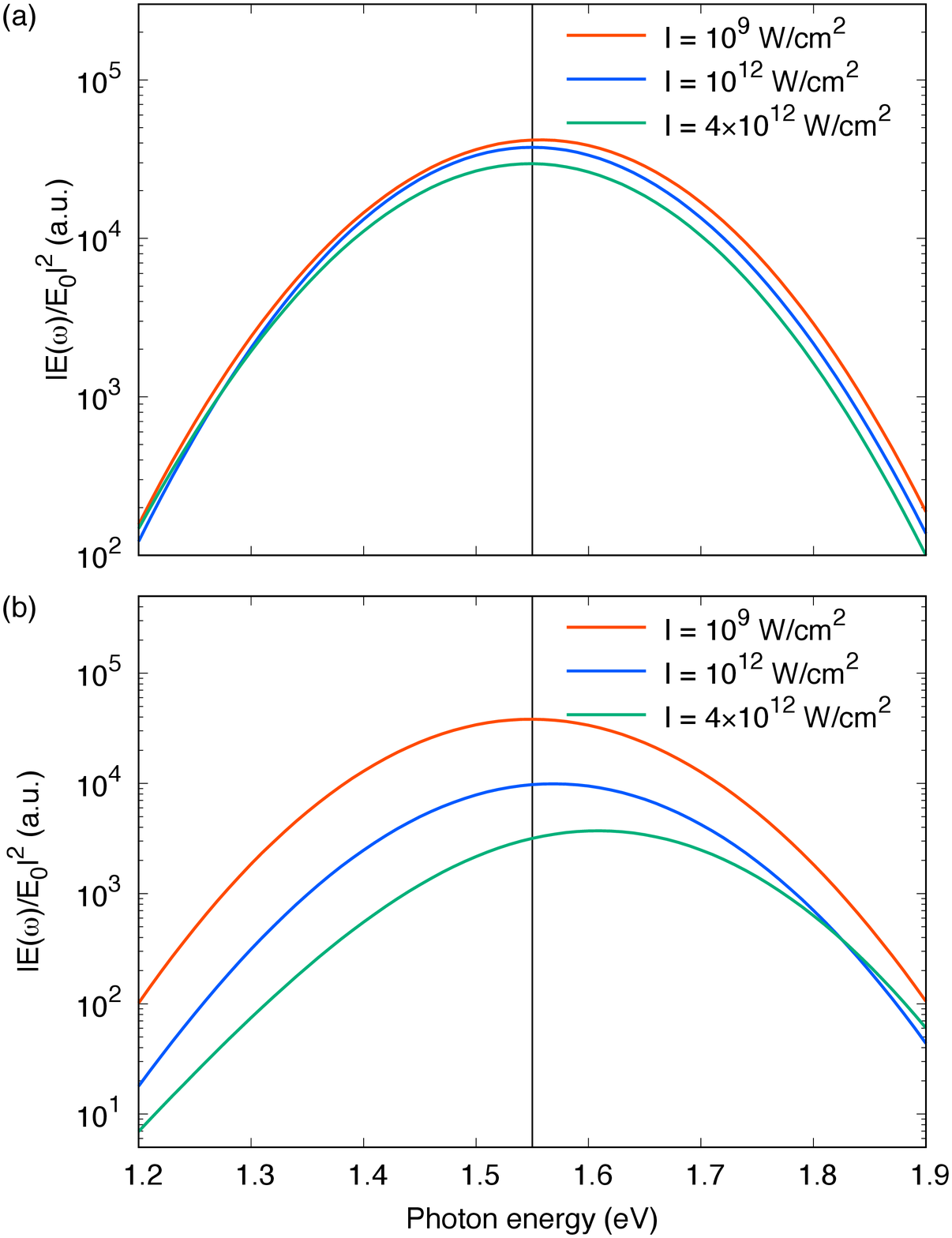}
    \caption{\label{fig:fundamental} 
    (a) Spectra of the  reflected fundamental waves normalized by the peak intensity of the incident pulse.
    The intensity is set to $I= 10^{9}$, $10^{12}$ , or  $4 \times 10^{12}$ W/cm${}^2$. 
    (b) The same as (a) but for the transmitted fundamental waves.
    }
\end{figure}

To verify the different mechanisms of THHG below and above 20 eV
described above, we revisit RHHG and THHG spectra emitted from the
1000~nm Si thin film that is shown again in
Fig.~\ref{fig:hhg_d1000}(a) by red and blue solid lines, respectively.
As we found in Fig.~\ref{fig:pulse_inv}(d), the THHG around 30 eV is
produced at the front surface and propagates through the film.  Then
we may expect that such THHG spectrum is proportional to the RHHG
spectrum multiplied by the penetration factor $\exp[-\alpha(\omega)d]$
where $\alpha(\omega)=(2\omega/c){\rm Im}n(\omega)$ is the attenuation
coefficient calculated from the dielectric function of Si.  This is
plotted by green dotted line in Fig.~\ref{fig:hhg_d1000}(a) for which
a constant factor is multiplied to roughly coincide with the THHG
spectrum around 30 eV.  The attenuation coefficient used in the plot
is shown in Figure~\ref{fig:hhg_d1000}(b) that is calculated by linear
response TDDFT.  Above 20 eV, the dielectric function of Si becomes
rather transparent and the penetration factor is close to unity.  In
the frequency region from 3 to 20 eV, the green dotted line in
Fig.~\ref{fig:hhg_d1000}(a) becomes extremely small due to the
penetration factor and cannot explain the THHG spectrum.  From these
observations, we can support the hypothesis that THHG above 20 eV is
indeed generated from the front surface.

To confirm that THHG below 20 eV is generated at the back surface, we
carry out the following analysis.  We first examine the electric field
at the front and the back surfaces in frequency domain around the
fundamental frequency.  In Fig.~\ref{fig:fundamental}(a), spectra of
electric fields at the front surface are compared for three different
intensities of the incident pulse, $I= 10^{9}$ W/cm${}^2$ (red line),
$I= 10^{12}$ W/cm${}^2$ (blue line), and $I= 4 \times 10^{12}$
W/cm${}^2$ (green line).  They are shown with a normalization such
that the incident pulse has the same amplitude.  In
Fig.~\ref{fig:fundamental}(b), the same spectra but at the back
surface are shown.  We observe the following facts.  For the electric
field at the front surface, the peak is equal to the average frequency
of the incident pulse, $\hbar\omega = 1.55$ eV, irrespective of the
intensity.  For the electric field at the back surface, the peak
position does not change for the weak intensity case of $I=10^{9}$
W/cm${}^2$.  However, as the intensity increases, the peak is
blue-shifted.  At the intensity $I= 4 \times 10^{12}$ W/cm${}^2$, the
peak position becomes $\hbar \omega=1.62$ eV.  We thus find that there
is a nonlinear effect that blue-shifts the peak frequency during the
propagation.  
As we noted previously in Fig.~\ref{fig:hhg}(b), THHG below 20 eV
shows blue-shift in the frequency as the intensity increases.  This
blue-shift can be naturally understood if we consider that the THHG
below 20 eV is generated at the back surface.

To further assure this mechanism, we carry out the following analysis:
We first extract the fundamental wave component from the electric field at the back surface, removing the HHG component.
For this purpose, we make the inverse-Fourier transformation of Eq.~(\ref{eq:inverse}) using the following window function,
\begin{eqnarray}
    w(\omega)= 
    \begin{cases}
    1 & (\omega < \omega_1), \\
    w^{}_{\rm Blackman} \left( \frac{\omega-\omega_1}{\omega_2-\omega_1} \right) & (\omega_1 < \omega < \omega_2), \\
    0 & (\omega_2 < \omega).
    \end{cases}
    \label{eq:fundamental}
\end{eqnarray}
with $\hbar \omega_1=3$ eV, and $\hbar \omega_2=5$ eV.  Using the
electric field thus obtained, we perform a single unit cell
calculation described in Sec.~\ref{sec:method}A.  We show the
calculated spectrum multiplied by a constant as the orange dotted line
in Fig.~\ref{fig:hhg_d1000}(a).  As seen in the figure, THHG (blue
line) and the orange dotted line agree well with each other, not only
in relative intensity of each order but also in positions of the peak
frequency.  Since the electric field at the back surface is rather
weak, it cannot produce harmonics above 20 eV.  From these analyses,
we conclude that THHGs below and above 20 eV have different origins
and the dip at 20 eV appears due to the combination of two mechanisms.

After gaining a detailed understanding of the mechanism of THHG, we
revisit thickness dependence of THHG shown in
Fig.~\ref{fig:thickness}(b).  As we mentioned previously, THHG signals
decrease as the thickness increases from 200 nm to 3000 nm.  A closer
look at the figure shows that the decreasing behavior of the signal
depends on the order.  While the THHG signal of 19th order decreases
exponentially (linearly in logarithmic plot), signals of other orders
behave differently.  Such difference of the signals can be understood
by the different mechanisms of THHG discussed previously.  Since THHG
below 20 eV is generated at the back surface, the intensities of the
3rd and 7th harmonics are expected to behave as the 3rd and 7th power of
the intensity of the fundamental wave at the back surface.  On the
other hand, the 19th HHG signal, whose frequency is about 30eV, is
generated at the front surface and its thickness dependence is
expected to be determined by the penetration factor,
$\exp[-\alpha(\omega)d]$.  These explanations are consistent with the
results shown in Fig.~\ref{fig:thickness}(b).  For the 13th harmonic  that
corresponds to about the dip position of the spectrum at 20 eV, the
THHG signal exhibits complex dependencies reflecting the competition
between the two mechanisms.

\section{Summary \label{sec:summary}}

We have investigated the impact of propagation effects on high-order
harmonic generation (HHG) in reflection and transmission pulses (RHHG
and THHG) from silicon (Si) thin films.  Using the multiscale
Maxwell-TDDFT method, we calculated the spectrum of RHHG and THHG for
Si films of various thickness up to 3000 nm and explored physical
mechanisms that determines their behavior.  The fundamental frequency
of the incident pulse is set to 1.55~eV, well below the direct bandgap
of Si.  We have found the  three following mechanisms to play an
important role: (1) Propagation effect on the strong pulse in the
frequency region of the incident pulse.  (2) Generation of HHG that
predominantly takes place at either the front or the back surfaces.  (3)
Propagation effect on the generated high harmonic waves under a linear
light-matter interaction.

We first note that HHG takes place efficiently at the front surface
since HHG is quite sensitive to the amplitude of the electric field
and it attenuates quickly as it propagate inside the thin film.  RHHG
is considered to be produced at the front surface.  We reported
previously that the intensity of RHHG depends sensitively on the
thickness of the film for thickness less than 200~nm, because of the
linear interference effect inside the thin film.  In the present work,
we have found that the interference effect becomes less important for
films above 200~nm, since the propagating wave inside the film
attenuates quickly.

Propagation effects appear in more complex way in THHG.  It shows
different behavior in the frequency regions below and above 20~eV.
This distinction comes from the linear absorption property of Si,
strongly absorptive below and almost transparent above 20~eV.  THHG
signals appear in the transmitted wave after the propagation inside
the thin film.  Then THHG that produced at the front surface attenuates
quickly for components below 20~eV.  It was found that THHG in the
frequency region below 20~eV is produced at the back surface.  It is
also observed that THHG below 20~eV shows blue-shift in the peak
positions.  This blue-shift can be understood in terms of the
nonlinear propagation effect of the pulse in the fundamental frequency
region.

The present results show that the propagation dynamics causes
interesting and significant effects in HHG from Si films of nano- to
micro-meter thickness and that the multiscale Maxwell-TDDFT scheme
provides a reliable description for them.  Since the method can be
applicable for three-dimensional nano-materials \cite{Uemoto2019-2}, it
is expected to be useful to design and simulate optimal
nano-materials that work as efficient HHG sources.

\begin{acknowledgements}
This research was supported by JST-CREST under grant number JP-MJCR16N5, and by MEXT Quantum Leap Flagship Program (MEXT Q-LEAP) Grant Number JPMXS0118068681, and by JSPS KAKENHI Grant Number 20H2649. Calculations are carried out at Fugaku supercomputer under the support through the HPCI System Research Project (Project ID: hp220120), and Multidisciplinary Cooperative Research Program in CCS, University of Tsukuba.
\end{acknowledgements}

\appendix


\end{document}